\documentclass[10pt,onecolumn]{article}
\usepackage[T1]{fontenc}
\usepackage[utf8]{inputenc}
\usepackage[margin=0.85in,columnsep=0.28in]{geometry}
\usepackage{graphicx}
\usepackage{times}
\usepackage[hidelinks]{hyperref}
\usepackage{titlesec}
\usepackage{caption}
\newcommand{\doi}[1]{\,\href{https://doi.org/#1}{doi:#1}}
\titlespacing*{\section}{0pt}{1.05ex}{0.45ex}
\titleformat{\section}{\normalfont\large\bfseries}{\thesection}{0.6em}{}
\title{\vspace{-3.5ex}\textbf{What Are We Measuring? Bonding, Trust, and the Evaluation of Human–Robot Relationships [Pre-print]}\vspace{-1ex}}
\date{Accepted at H-STAR Workshop (https://hstar.qmic.com/), RO-MAN 2026}

\author{Imran Khan\\[2pt]
\small Department of Computer Science, University of Warwick, Coventry, UK\\
\small \texttt{imran.khan.3@warwick.ac.uk}}

\begin{document}
\maketitle
\thispagestyle{empty}
\begin{abstract}
\noindent
In human--robot interaction, relationship quality is often quantified using self-report measures, particularly related to ``trust'', such that a robot's trustworthiness comes to serve as an index of how close or ``bonded'' a human feels to it. I argue that this is a category error: trust and social bonding are distinct constructs, differing in their antecedents, their timescales, their bodily signatures, the human experience they produce, the robot responses they call for, and the ethical concerns they raise. I propose that we view them as independent dimensions, and describe the resulting two-dimensional space of possible relationship states under this view, with four configurations: \textit{avoidance, functional, dependence, }and \textit{symbiosis.} I then draw out some consequences for human-state-aware robotics: (1) social bonding is an explicit estimation target distinct from trust, (2) it should condition online adaptation (3) it reframes what a ``failure'' means, and (4)  it raises the possibility of identifying dysfunctional relationships, in which a user remains attached to a robot that no longer merits reliance. This is ongoing work, offered in part to prompt the field to reconsider what it means to evaluate relationship quality in human--robot dyads.
\end{abstract}
\section{Introduction}
In studies of long-term human--robot interaction, trust is commonly treated as a summary measure of relationship quality~\cite{campagna2025}. A high trust score is read as a healthy relationship (or the potential for one), and a low one as a deteriorating relationship; trust is increasingly measured, most often with self-report questionnaires, and used to drive adaptive robot behaviour~\cite{hancock2011,mayer1995}. For short, task-oriented interactions i.e. data collection settings, this is reasonable. For the long-term robot deployments that much work in the field aspires to, short-term trust measurements fall short.

The issue is not the accuracy of ``trust'' measurements, per se, but what it actually tells us about the dynamics between social agents. Trust is only part of what social relationships involve. In human relationships, the two routinely come apart: a parent's bond with their child does not rest on how trustworthy the child is---it holds toward an infant who can do nothing to earn it, and endures toward an adult child one has learned not to depend on. The same dissociation appears in human--robot dyads. Consider an older adult who becomes distressed when a companion robot is removed despite never having relied on it for any task~\cite{paro}, and who therefore has no functional reason to ``trust'' it. Described only on a single dimension (trust), such a relationship is either missed or misclassified. The ongoing work presented here argues that relationship quality in HRI is better represented by two independent constructs, social bonding and trust. I propose that estimating this two-dimensional state, and understanding the consequences of doing so, should be a consideration in adaptive, human-state-aware robotics. It requires, amongst other things, delineating between these two constructs and understanding how each is produced, how it is expressed---in what a person reports, but also in their behaviour and physiology---and what response each warrants from the robot. In what follows, I make this distinction precise, describe the space of relationship states it generates, and draw out its consequences for estimation, adaptation, and the ethics of the bonds robots invite.

\section{Distinguishing Bond from Trust}
Outside HRI, the distinction between trust and social bonding is well established, and so this is not a novel consideration in itself. Trust, as operationalised in most HRI research, is a cognitive appraisal of whether a robot will perform reliably~\cite{lewis1985,mcallister1995}. It is forward-looking, updates quickly in response to the robot's behaviour, and is fragile. In other words, a single salient failure can substantially reduce trust in a robot~\cite{hancock2011}. The trust literature does distinguish an affect-based form of trust from this cognitive one, but even affect-based trust is an appraisal of a partner, rather than a physiological tie to them.

Social bonding is a different process. Although somewhat-contentious, attachment theory describes social bonding in terms of proximity maintenance, use of the partner as a safe haven and secure base, and distress on separation~\cite{bowlby1969,ainsworth1978}. A bond accumulates gradually through repeated positive interaction, rather than updating on discrete performance events. In the neuroscience of pair bonding, it is associated with oxytocinergic signalling that develops over extended periods, attenuates amygdala reactivity, and reduces cortisol~\cite{bosch2018,feldman2012}. Its signatures are largely involuntary and bodily: lower physiological arousal in the partner's presence, reduced interpersonal distance, and behavioural synchrony~\cite{khan2024}. Where trust reflects what a robot \textit{does}, bonding reflects what it is \textit{like} to be with the robot. These two need not covary.

The two constructs also play different roles in a relationship. Trust functions as a decision variable: it governs whether a person relies on the robot's outputs at a given moment, and because it is event-driven, it tracks recent performance closely. A bond functions instead as a stabiliser of the relationship over time, and it is through this stabilising role that an established bond can absorb the effect of a robot's failures. The reasoning is twofold. First, the physiology of bonding blunts the response to the failure itself: the oxytocinergic social buffering mechanisms that lower stress in a bonded partner's presence also attenuate amygdala threat reactivity and cortisol release~\cite{bosch2018,feldman2012,khan2024}, so an error from a bonded robot is appraised as less threatening than the same error from a stranger. Second, a bond supplies an interpretive context: because it integrates a long history of positive interaction, a single failure is weighed against that history rather than treated as diagnostic, and is more readily attributed to circumstance than to the robot's character, much as accommodation in close human relationships dampens the impact of a partner's transgressions~\cite{mikulincer1998}. In neither case does the bond change what the robot did; it changes how much that failure moves the relationship, which is why the same error can be near-terminal in a new dyad yet survivable in a bonded one.

That they are separable is evident in ordinary cases. A person may trust a machine that does its job well, like a robot vacuum cleaner, without feeling any attachment to it, or may remain attached to a romantic partner they have ceased to trust. The latter is the pattern attachment theory identifies in anxious attachment, where a strong bond persists toward a chronically unreliable other~\cite{mikulincer1998}. Neither of these states, however, would be correctly represented using traditional trust-based measurements.

\section{Trust as a (Failed) Proxy for Relationship Quality}

In many HRI studies, relationship quality has been quantified by self-report questionnaires, measuring trust~\cite{schaefer2016,ullman2019,jian2000} amongst other things~\cite{bartneck2009}, with self-report remaining the dominant assessment method in the field~\cite{campagna2025}. 

Validated trust instruments such as the Trust Perception Scale-HRI~\cite{schaefer2016}, the Multi-Dimensional Measure of Trust~\cite{ullman2019}, and the Trust in Automated Systems survey~\cite{jian2000}, alongside broader perception scales such as the Godspeed series~\cite{bartneck2009} and numerous study-specific Likert scales, are a common means of assessing how a relationship is developing (at least in a unidirectional sense, i.e. from the human to robot). Because these instruments are in routine use, and no comparably established instrument targets social bonding, a rising trust score is used as evidence that the relationship is developing. Trust is thereby used as a proxy for the relationship as a whole.

The substitution persists even where bonding is the explicit object of study. A recent review of human--AI relationships notes that work intending to assess attachment or companionship commonly falls back on items concerning trust, competence, or positive affect as proxies for attachment~\cite{attachai2025}: the questionnaire labelled ``attachment'' is often a trust questionnaire under another name.

When a genuine ``bond'' signal is recorded, it tends to be absorbed into the trust construct rather than treated separately. Miller et al.~\cite{miller2021} had a domestic robot approach participants and measured comfortable interpersonal distance alongside trust; comfort distance decreased as learned trust increased, and the two were reported as aspects of a single trust process. However, if we consider these two to be separate constructs, this approach takes a bond signal (narrowing distance) and logs it onto a trust instrument. Such a design cannot detect the case in which the two diverge, with proximity remaining close while trust falls.

The reliance on trust instruments also breaks down where there is no task to be reliable at, i.e. no reason for trust to form in the first place. In long-term care studies using PARO~\cite{paro}, residents talk to the robot as though it were a person and show distress when it is withdrawn, yet a trust questionnaire has little to measure, because the robot performs no function whose reliability it would assess. An affiliative relationship exists, but goes unquantified, yet it is precisely the relationship the field increasingly aims to support.

An alternative exists but has not been paired with trust measurement. Biobehavioural proxies of bonding, grounded in the physiology of social bond formation, have been proposed for HRI: reduced physiological stress in the robot's presence (social buffering), narrowing of interpersonal distance, and inter-dyad behavioural synchrony~\cite{khan2024}. To my knowledge, no longitudinal study to date records such bond proxies together with a validated trust instrument.

\section{A Two-Axis Model of Relationship State}
\begin{figure}[t]
\centering
\includegraphics[width=0.96\linewidth]{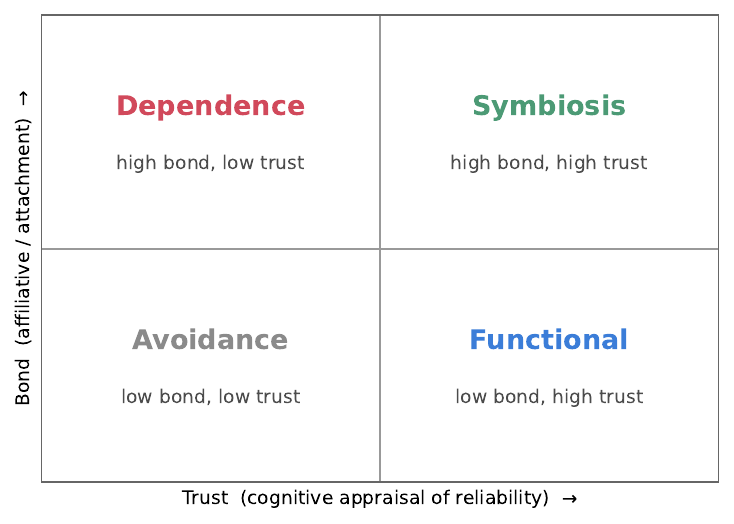}
\caption{The conceptual bond\,$\times$\,trust space, with the two dimensions shown as separate axes. Further descriptions of each quadrant are given in this section}
\label{fig:quad}
\end{figure}

If we agree to treat bond and trust as independent, we can then consider the state of a dyad at a given time as a point in a two-dimensional space. We visualise this space (which I simply call the bond\,$\times$\,trust space) in Fig.~\ref{fig:quad}, showing the two respective dimensions as separate axes. This two-dimensional space can further be broken down into four distinct quadrants, which we can call \emph{avoidance} (low bond, low trust), \emph{functional} (low bond, high trust), \emph{dependence} (high bond, low trust), and \emph{symbiosis} (high bond, high trust). Each comes with a distinct set of behavioural signatures, potential risks, and implications for how a robot should respond. We describe these quadrants in more detail as follows.

\textbf{Avoidance (low bond, low trust).} Here, no relationship has formed. This is also the initial state of any dyad, and the state to which it returns when early failures reduce trust before a bond can develop. This was the case in ~\cite{esterwood2023} where,  across 240 participants working with a robot co-worker, trustworthiness was not fully restored after repeated violations, and none of the repair strategies tested (apology, denial, explanation, promise) returned it to baseline. Without any established bond present to absorb violations of trust, failures accumulate, and the dyad remains here.

\textbf{Functional (low bond, high trust).} The robot is trusted as a competent performer, but no meaningful affiliative relationship (social bond) exists yet. This is the implicit (and likely unintended) target of many HRI studies, and is appropriate for short, task-focused interaction. For instance, in ~\cite{robinette2016}, all 26 participants followed a robot during a simulated emergency, although half had seen it perform poorly minutes earlier. Situational trust had been established with no interaction history behind it, producing reliance that the robot's demonstrated performance did not warrant. Over longer horizons, however, this state risks being fragile. With no social bond present, a serious failure (or repeated set of failures) has the potential to affect trust seriously, and a relationship that appeared healthy can quickly deteriorate.  

\textbf{Dependence (high bond, low trust).} A social bond has been established, but functional trust is low. This is both the region a trust measure alone cannot represent and the one with the greatest potential for harm. I argue that this is far from a marginal case, but rather a likely endpoint of long deployments in which social quality appears to be maintained as task performance declines. In~\cite{robotstayed}, four years after an in-home literacy deployment, 18 of 19 families had retained a reading-companion robot their children had outgrown, describing it as a member of the household and treating its removal as a loss: the bond had simply outlived the robot's usefulness. The harmful version is the same configuration at higher stakes: a vulnerable user, such as an older adult in care, who keeps relying on a robot for a task it can no longer perform e.g. taking its reminders, guidance, or assurances at face value, because of the social bond that remains. The dependence is actively sustained by the bond, and a trust-only account, which records such a dyad as no relationship at all, cannot identify this potentially harmful relationship.

\textbf{Symbiosis (high bond, high trust).} Here, strong social bonds along with high levels of trust have been established. This state is more robust than either dimension alone: as set out in Section~2, an established bond buffers trust against the effect of occasional failures, both by dampening the physiological response to them and by supplying a history against which they are read. It is, I argue, the idealised target state for any long-term human--robot dyad: the robot is both relied upon and cared for, and the relationship can withstand the inevitable failures without collapsing. Kanda et al.'s two-month school field trial~\cite{kanda2007} is consistent with the trajectory toward it: children who interacted with the robot more frequently reported stronger relationships over time. No study, however, has deliberately targeted this state, which would require tracking both dimensions simultaneously.

\section{How Does a Bond x Trust Framework Affect Human-State-Aware Robotics?}
Why should we delineate these configurations of human-robot relationship states in this way? In the spirit of this workshop, we should consider what this framework can contribute to human-state-aware robotics. I propose that this separation reframes three things that a human-state-aware system must do, and adds an ethical constraint on all of them.

\textbf{Bonded state is an estimation target in its own right.} Human-state estimation in HRI has largely concerned momentary states such as affect~\cite{markelius}, engagement~\cite{ravandi}, cognitive load~\cite{ravandi2}, and, where the relationship is addressed, trust~\cite{chen2020}. The two-axis framework argues for explicit measurements of whether (and how strongly) a human is bonded to the robot. I have argued that this is not recoverable from a trust questionnaire because bonding and trust can, and do, dissociate. It therefore calls for understanding and measuring the multimodal, largely involuntary signals through which bonds express themselves: reduced physiological arousal in the robot's presence, narrowing interpersonal distance, and behavioural synchrony~\cite{khan2024}. The target is not only how much a human trusts the robot to fulfil its role, but whether they care about it, and thus the two must necessarily be read from different channels.

\textbf{Robot adaptation should be conditioned on both axes.} A robot that is designed to adapt to a single trust estimate cannot behave or transition appropriately across the four regions in the bond\,$\times$\,trust space, since these regions call for different responses. For instance, in the functional region, the priority should be to facilitate bond formation, since a bond-free relationship, however high its trust score, is fragile~\cite{esterwood2023} and susceptible to collapse with repeated failures. The dependence region is more delicate. Adaptive robots are typically tuned to maximise engagement, treating proximity, positive affect, and low arousal as signs that the interaction is going well; but in this region those signals are produced by the bond rather than by the robot's performance, which has declined, so a policy that optimises them simply deepens the user's reliance on a robot that no longer merits it. The appropriate response runs against these signals: rather than sustaining engagement, the robot should work to recalibrate the user's reliance to the robot's actual performance, for instance by making its failures and uncertainty legible or, for vulnerable users, by escalating to a carer.

\textbf{What ``failure'' means depends on the relational state.} Trust-based accounts treat a failure as a fixed decrement to be repaired. On the two-axis view, and for the reasons given in Section~2, its meaning is relational: the same error is a near-terminal event in a bond-free dyad, where repeated violations leave trust persistently depressed~\cite{esterwood2023}, but a survivable one in a bonded dyad, where months-long deployments tolerate imperfect robots and deepen into attachment~\cite{kanda2007,robotstayed}. A system that estimates the relational state can therefore interpret and respond to its own failures according to the bond it has, rather than applying a single repair policy.

\textbf{Bonded states raise ethical questions that trust does not.} I highlight the dependence region as not only a measurement blind spot but an ethical one. A user who remains attached to a robot whose performance has declined continues to depend on it because of a strong social bond; in vulnerable populations, such as older adults with dementia or children, this dependence may be neither noticed nor consented to. A retained but obsolete companion robot~\cite{robotstayed} is a benign instance, but one can also imagine a dependent, vulnerable user paired with an unreliable robot with which they have formed a dysfunctional bond. Making bond an explicit, estimated state forces the question into the open. A system that can detect a dysfunctional bonded state (i.e. a state of \textbf{dependence}) must also decide what it owes the user in that state, and a field that builds robots people bond to takes on responsibility for the dependence it creates. Surfacing this state, rather than leaving it hidden inside a trust score, is the minimum required to address it.

\section{Conclusion}
Relationship quality in human--robot interaction is not adequately captured by trust alone, yet it is often through trust-based measures that we assess it. I have argued that trust and social bonding are distinct constructs, with different antecedents, timescales, behavioural signatures, and functional roles, and that the cognitive and neuroscientific literature has treated them as such for decades. Treating them as two dimensions is a modest conceptual move, but it has direct consequences for human-state-aware robotics. Whether a human is ``bonded'' to a robot becomes an estimation target distinct from trust, that this relational state should condition online adaptation, the notion of ``failure'' depends on the relational state that it occurs within, and it exposes ethical questions about dysfunctional bonded states that might otherwise be latent. With this in mind, we, as a field, should reconsider what it means to evaluate relationship quality in human--robot dyads, and begin to build the measures that such an evaluation would require.

\end{document}